\let\includefigures=\iftrue
%
\let\useblackboard=\iftrue
%
%
\newfam\black
\input harvmac.tex
\input epsf
\includefigures
\message{If you do not have epsf.tex (to include figures),}
\message{change the option at the top of the tex file.}
\def\figin{\epsfcheck\figin}\def\figins{\epsfcheck\figins}
\def\epsfcheck{\ifx\epsfbox\UnDeFiNeD
\message{(NO epsf.tex, FIGURES WILL BE IGNORED)}
\gdef\figin##1{\vskip2in}\gdef\figins##1{\hskip.5in}
\else\message{(FIGURES WILL BE INCLUDED)}%
\gdef\figin##1{##1}\gdef\figins##1{##1}\fi}
\def\DefWarn#1{}
\def\figinsert{\goodbreak\midinsert}
\def\ifig#1#2#3{\DefWarn#1\xdef#1{fig.~\the\figno}
\writedef{#1\leftbracket fig.\noexpand~\the\figno}%
\figinsert\figin{\centerline{#3}}\medskip\centerline{\vbox{\baselineskip12pt
\advance\hsize by -1truein\noindent\footnotefont{\bf Fig.~\the\figno:} #2}}
\bigskip\endinsert\global\advance\figno by1}
\else
\def\ifig#1#2#3{\xdef#1{fig.~\the\figno}
\writedef{#1\leftbracket fig.\noexpand~\the\figno}%
\global\advance\figno by1}
\fi
\useblackboard
\message{If you do not have msbm (blackboard bold) fonts,}
\message{change the option at the top of the tex file.}
\font\blackboard=msbm10 scaled \magstep1
\font\blackboards=msbm7
\font\blackboardss=msbm5
\textfont\black=\blackboard
\scriptfont\black=\blackboards
\scriptscriptfont\black=\blackboardss
\def\Bbb#1{{\fam\black\relax#1}}
\else
\def\Bbb{\bf}
\fi
%
\def\yboxit#1#2{\vbox{\hrule height #1 \hbox{\vrule width #1
\vbox{#2}\vrule width #1 }\hrule height #1 }}
\def\fillbox#1{\hbox to #1{\vbox to #1{\vfil}\hfil}}
\def\ybox{{\lower 1.3pt \yboxit{0.4pt}{\fillbox{8pt}}\hskip-0.2pt}}

\def\comments#1{}

\def\QR{\Bbb{R}}

\def\II{\relax{I\kern-.10em I}}

\def\hk{hyperk\"ahler\  }

\def\IZ{\relax\ifmmode\mathchoice
{\hbox{\cmss Z\kern-.4em Z}}{\hbox{\cmss Z\kern-.4em Z}}
{\lower.9pt\hbox{\cmsss Z\kern-.4em Z}}
{\lower1.2pt\hbox{\cmsss Z\kern-.4em Z}}\else{\cmss Z\kern-.4em
Z}\fi}
\def\IB{\relax{\rm I\kern-.18em B}}
\def\IC{{\relax\hbox{$\inbar\kern-.3em{\rm C}$}}}
\def\ID{\relax{\rm I\kern-.18em D}}
\def\IE{\relax{\rm I\kern-.18em E}}
\def\IF{\relax{\rm I\kern-.18em F}}
\def\IG{\relax\hbox{$\inbar\kern-.3em{\rm G}$}}
\def\IGa{\relax\hbox{${\rm I}\kern-.18em\Gamma$}}
\def\IH{\relax{\rm I\kern-.18em H}}
\def\II{\relax{\rm I\kern-.18em I}}
\def\IK{\relax{\rm I\kern-.18em K}}
\def\IN{\relax{\rm I\kern-.18em N}}
\def\IP{\relax{\rm I\kern-.18em P}}

%
\def\inbar{\,\vrule height1.5ex width.4pt depth0pt}

\font\cmss=cmss10 \font\cmsss=cmss10 at 7pt
\def\IR{\relax{\rm I\kern-.18em R}}

%


%

\def\gs{g_s}
\def\lp10{l_P^{10}}
\def\lp11{l_P^{11}}
\def\R11{R_{11}}

\Title{\vbox{\baselineskip12pt\hbox{hep-th/0007255}
\hbox{RUNHETC-2000-07}}}
{\vbox{
\centerline{Noncommutative Scalar Solitons at Finite $\theta$}}}
\smallskip
\centerline{Chen-Gang Zhou}
\medskip
\centerline{Department of Physics and Astronomy}
\centerline{Rutgers University }
\centerline{Piscataway, NJ 08855--0849}
\medskip
\centerline{\tt czhou@physics.rutgers.edu.}
\bigskip
\noindent

We investigate the behavior of the noncommutative scalar soliton
solutions at finite noncommutative scale $\theta$. A detailed analysis
of the equation of the motion indicates that fewer and fewer soliton
solutions exist as $\theta$ is decreased and thus the solitonic sector
of the theory exhibits an overall hierarchy structure. If the
potential is bounded below, there is a finite  $\theta_c$ below which
all the solitons cease to exist even though the noncommutativity is
still present. If the potential is not bounded below, for any nonzero
$\theta$ there is always a soliton solution, which becomes singular
only at $\theta = 0$. The $\phi^4$ potential is studied in detail and
it is found the critical $(\theta m^2)_c =13.92$ ($m^2$ is the
coefficient of the quadratic term in the potential) is universal for
all the symmetric $\phi^4$ potential.

\Date{July 2000}
%
\nref\cds{A.Connes, M.R.Douglas and A.Schwarz," Noncommutative Geometry and
Matrix Theory:compactification on Tori", JHEP 9802 (1998) 003,
hep-th/9711162.}
\nref\sw{N.Seiberg and E.Witten, " String Theory and Noncommutative
 Geometry",JHEP 9909 (1999) 032, hep-th/9908142.}
\nref\w{E. Witten, ``Noncommutative Tachyons and String Field Theory'', hep-th/0006071.}
\nref\filk{T.Filk, ``Divergencies in a Field Theory on Quantum Space'', Phys.Lett.B376 (1996) 53.}
\nref\bs{ D.Bigatti and L.Susskind, ``Magnetic Fields, Branes and Noncommutative Geometry'', hep-th/9908056.}
\nref\chr{I.Chepelev and R.Roiban, ``Renormalization of Quantum Field Theories on Noncommutative R**d. I: Scalars'', hep-th/9911098.}
\nref\mrs{S.Minwalla, M.V.Raamsdonk and N.Seiberg, ``Noncommutative Perturbative Dynamics'', hep-th/9912072.}
\nref\rs{M.V.Raamsdonk and N.Seiberg, ``Comments on Noncommutative Perturbative Dynamics'', hep-th/0002186.}
\nref\ggrs{H.O.Girotti, M.Gomes, V.O.Rivelles and A.J.da Silva, ``A Consistent Noncommutative Field Fheory: the Wess-Zumino Model'', hep-th/0005272.}
\nref\gs{S.S.Gubser, S.L.Sondhi, ``Phase Structure of Noncommutative Scalar Field Theories'', hep-th/0006119.}
\nref\gms{R.Gopakumar, S.Minwalla and A.Strominger, "Noncommutative Solitons",
hep-th/0003160.}
\nref\dmr{K.Dasgupta, S.Mukhi, G. Rajesh, ``Noncommutative Tachyons'', hep-th/0005006.}
\nref\hklm{J.A.Harvey, P.kraus, F.Larsen, E.J.Martinec, ``D-branes and Strings ss Noncommutative Solitons'', hep-th/0005031.}
\nref\sena{A.Sen, ``Descent relations among bosonic D-branes'', Int.J.Mod.Phys.A14(1999) 4061, hep-th/9902105.}
\nref\senb{A.Sen, ``Universality of the Tachyon Potential'', JHEP 9912 (1999) 027, hep-th/9911116.}
\nref\sz{A. Sen and B.Zwiebach,  ``Tachyon Condensation in String Field Theory'', hep-th/9912249. }
\nref\hk{J.A.Harvey and P.Kraus, ``D-branes as Lumps in Bosonic Open String Field Theory'',JHEP 0004:012, 2000 hep-th/0002117.}
\nref\kjmt{R.de Mello Koch, A.Jevicki, M.Mihailescu and R.Tatar, ``Lumps and P-branes in Open String Field Theory'', hep-th/0003031.}
\nref\ks{V.A.Kostelecky and S.Samuel, ``On a Nonperturbative Vacuum for the Open Bosonic String'', Nucl.Phys.B336(1990) 263.}
\nref\bsz{N.Berkovits, A.Sen and B.Zwiebach, ``Tachyon Condensation in Superstring Field Theory'', hep-th/0002211.}
\nref\sochi{C. Sochichiu, ``Noncommutative Tachyonic Solitions. Intercation with Gaube Field'', hep-th/0007217.}
\nref\gmsA{R.Gopakumar, S.Minwalla and A.Strominger, "Symmetry Restoration and Tachyon Condensation in Open String Theory'', hep-th/0007226.}


\newsec{Introduction}

Recently it was found that noncommutative geometry arises naturally in
string theory with a constant $B$ field background \refs{\cds,
\sw}. In particular, in the large $B$ limit, the string field algebra
factors into a direct product with the noncommutative algebra being an
independent subalgebra\refs{\w}, resulting in a noncommutative field
theory as a decoupled low energy effective description of string
theory in this limit. It is highly nonlocal and contains infinitely
high order derivative terms, but in a controlled and self-consistent
way. Compared to the commutative case, the renormalizability is
improved but remains an open question, while nontrivial behavior such
as UV/IR mixing adds to the difficulties
\refs{\filk,\bs,\chr,\mrs,\rs,\ggrs}. The phase structure of the
noncommutative scalar field theory is analyzed in the $\phi^4$ case
\refs{\gs} in which an unusual phase structure is uncovered.

The soliton sector of the noncommutative scalar field theory also
exhibits an intriguingly rich structure. In the limit of large
noncommutative parameter $\theta$ and ignoring the kinetic term, the
solitons can be explicitly constructed via an isomorphism between the
noncommutative fields and the operators on a single particle Hilbert
space \refs{\gms}, and it was found that there are infinitely many
solutions as long as the potential has more than one extremum. In
spacetime, the soliton interpolates between the pseudo-vacuum at the
core and the true vacuum at spatial infinity. This construction has
been applied in the description of the unstable D-branes and of
tachyonic condensation \refs{\dmr, \hklm}, based on the idea that
D-branes can be constructed as solitons or lumps in the open string
field theory \refs{\sena,\senb,\sz,\hk,\kjmt,\ks,\bsz}. These solitons
are expected to disappear in the commutative limit $\theta=0$ because
Derrick's theorem states that there is no soliton solution in more
than 1+1 dimensions. Precisely how this happens, as $\theta$ changes
from infinity to zero, will be investigated in this paper.

The general picture we found is as follows. First, all the solutions
at $\theta =\infty$ can be extrapolated into the finite $\theta$
region, when the contribution from the kinetic term is taken into
account. As we will show later, the noncommutative parameter $\theta$
becomes an overall multiplicative factor of the scalar potential,
after a simple scaling. As $\theta$ decreases, the potential seen by
the soliton scales down, which finally makes the soliton solution
impossible. Each soliton at $\theta=\infty$ has its own critical point
in $\theta$, and when $\theta$ gets smaller this particular solution
disappears. In general different solutions at $\theta=\infty$ have
different critical points for their own existence, so there is a
hierarchy structure controlled by $\theta$, and we will see fewer and
fewer soliton solutions as $\theta$ is decreased.

Depending on whether the potential is bounded below we have
qualitatively different results. If the polynomial potential is
bounded below and assumed to have its global minimum at the origin,
there exists a lowest critical point $\theta_c$ which is nonzero. The
bound for this critical point can be estimated for each particular
potential, in some cases it can even be found precisely, such as in
$\phi^4$ theory, which will be studied in detail in this paper. If
instead the potential is not bounded below, such as for the cubic
potential, there always exists a certain soliton solution at any
nonzero $\theta$ and it becomes singular only in the commutative limit
$\theta =0$. Thus the noncommutative soliton owes its existence not
only to the noncommutativity, but also to the dynamics.

The paper is organized as follows. In section 2 we review the
construction of the noncommutative scalar soliton at
$\theta=\infty$. In particular the isomorphism between the
noncommutative algebra and the operator algebra on a single particle
Hilbert space is discussed in detail. Ignoring the kinetic energy, the
soliton in this limit is easily obtained by solving an algebraic
equation. In section 3, we consider the solitons at finite $\theta$
for a general polynomial potential, and provide a proof of our result
through a qualitative study of the full equation of motion expanded in
the projection operator basis. In section 4, we provide a detailed
numerical study of the soliton solution at finite $\theta$, for the
$\phi ^4$ theory. We use a more natural dimensionless parameter
$\theta m^2$ instead of $\theta$, where $m^2=V'(0)$. The approximation
method in \refs{\gms} gives a good estimation of $(\theta
m^2)_c$. Using a simple scaling argument, we find that the
dimensionless critical parameter $(\theta m^2)_c$ is the same for all
$\phi^4$ potential with  two degenerate vacua. In section 5, we
discuss string theoretic implications of our results and give our
conclusions.

\newsec{Review of Noncommutative Scalar Solitons}

The basis for noncommutative geometry lies in the deformation of the
usual commutative product of the smooth functions on a flat space
$\QR^n$, into the noncommutative star product. The simplest
realization of this deformation is Weyl quantization, when a Poisson
bracket structure over the space of smooth functions
$C^{\infty}(\QR^n)$ is determined by a constant bivector field
$\theta^{ij}{\partial \over {\partial x^i}}{\partial \over {\partial
x^j}}$, and the product assumes the form
\eqn\StarProd{ 
f(x)\ast  g(x)=e^{-i\theta^{ij}  {\partial  \over  {\partial a^i}}
{\partial \over {\partial b^j}}} f(x+a)g(x+b)|_{a=b=0}. }

Actually this is the familiar quantization procedure in quantum
mechanics, as the Poisson bracket is lifted up to the canonical
commutation relations, and functions on the phase space become
operators on the quantum Hilbert space. So after the deformation
quantization, the noncommutative star algebra $(C^{\infty}(\QR^n),
\ast)$ is naturally isomorphic to the operator algebra over a quantum
Hilbert space corresponding to a finite number of particles. Under the
Weyl prescription, the algebraic isomorphism is
\eqn\Weyla{
f(x)=  {1\over (2\pi)^n}  \int  d^{2n}k\tilde{f}(k)e^{-i(kx)}, }
\eqn\Weylb{
\hat{f}(\hat{x})={1\over{(2\pi)^n}}
\int d^{2n}k\tilde{f}(k)e^{-i(k\hat{x})} .
}

The star product is mapped to the operator product, and the
integration of the function over the phase space is equal to the trace
of the corresponding operator in the Hilbert space $\CH$
\eqn\IosProd{
f\ast  g \longleftrightarrow  \hat{f} \hat{g}, }
\eqn\Trace{
{1\over (2\pi)^n}\int d^{2n}xf(x)=Tr_{\CH}\hat{f}.}

As Hilbert space is separable and naturally equipped with a positive
definite inner product, it always has a complete set of orthonormal
basis of vectors $\{ |n \rangle,  n=0,1,2,... \}$. Then the bounded
linear operators on the Hilbert space have a corresponding basis
composed of operators $\{|m\rangle \langle n|,
m,n=0,1,2,...\}$. Conversely, the above isomorphism, \Weyla\ and
\Weylb, allows us to find the smooth function corresponding to each
operator.

We will give the detailed construction in the case that the space is
two dimensional, and the generaliation to the higher dimensional case
is easy. $\CH$ now is a single particle Hilbert space. Using
\StarProd, the complex variable $\bar{z}$ and $z$ are mapped to the
creation operator $a^{\dagger}$ and the annihilation operator $a$
respectively.  We choose the simple harmonic oscillator basis, which
are eigenstates of the operator $a^{\dagger}a\sim{r^2\over 2}$
\eqn\SHO{
|n\rangle={(a^{\dagger})^n\over {\sqrt{n!}}}|0\rangle, \qquad a|n
 \rangle =\sqrt{n}|n-1 \rangle ,
\qquad a^{\dagger}|n \rangle =\sqrt{n+1}|n+1 \rangle .
}    A real function $\phi^{\dagger}(x)=\phi(x)$ corresponds to an
hermitian operator, and so can be diagonalized using a unitary
operator $U$ ($U^{\dagger}U=UU^{\dagger}=I$)
\eqn\opphi{
\hat{\phi}=U(\sum_{n=0}^{\infty} \lambda_n |n \rangle \langle  n | ) U^{\dagger} ,
\qquad \lambda_n \in \QR .
}

The projection operator is easily expressed in the normal ordered
form,  $|n\rangle\langle n|={1\over
{n!}}{:}(a^{\dagger})^ne^{-a^{\dagger}a}a^n{:}$ which is proportional
to the n-th Laguerre polynomial $L_n$ in momentum space
\eqn\NormalInP{
 \phi_N^{(n)}(k)=e^{-{k^2\over  2}}L_n({k^2\over  2}).  } Actually
Weyl ordering and normal ordering are two equivalent isomorphisms from
the star algebra to the operators on Hilbert space (Weyl quantization
uses the symmetric ordering), and in particular they differ by an
integration kernel in momentum space
\eqn\ordering{
\tilde{f}_W(k) =\tilde{f}_N(k)e^{k^2\over 4 }.
}  Now, using the relation \Weylb \NormalInP\ and \ordering, we can
easily find that the projection operator corresponds to a radially
symmetric function
\eqn\Field{\eqalign{
\phi_n(r^2) & = {1\over {2\pi}}\int d^2ke^{-{k^2\over 4}}L_n(k^2/2)e^{-ik.x} \cr
& = 2(-1)^ne^{-r^2}L_n(2r^2) }. }

The functions $\phi_n(x)$ have the same properties under the star
product as the corresponding projection operators. This will greatly
facilitate the construction of the noncommutative solitons.

Consider noncommutative scalar field theory in 2+1 dimensions with
noncommutativity only in the spatial dimensions. The soliton is the
classical extremum of the energy functional
\eqn\energyA{ E[\phi]=\int
d^2x[{{\partial}_\mu \phi}*{{\partial}^\mu  \phi}+V(\phi)].  }

Upon changing variables to $(z, \bar{z})$, and performing a rescaling
$z \rightarrow {z \over \sqrt{\theta}}$, the star product will be
independent of $\theta$, and the sole effect of the noncommutative
parameter $\theta$ will appear as an overall scale factor for the
potential
\eqn\energyB{ E=\int d^2z({1\over 2}\partial
\phi\ast\bar{\partial}\phi+\theta V(\phi)). 
}    In the operator representation, the energy functional becomes
\eqn\energyC{
E(\phi)=K(\phi)+U(\phi),  \qquad  K(\phi)=Tr[a,\phi][\phi,a^+], \qquad
U(\phi)=\theta  TrV(\phi).   }

At $\theta=\infty$, the kinetic energy is much smaller than the
potential energy and so can be ignored. The potential $U(\hat{\phi})$
has a $U(\infty)$ symmetry, and the scalar field can be diagonalized
as in \opphi. In the operator representation, the potential is a
function of the coefficient series $\{\lambda_n, n=0, 1,...\}$, where
the $\lambda_n$'s are decoupled from each other
\eqn\pota{
E(\phi)=\theta V(\phi)=\theta\sum_{n=0}^{+\infty}V({\lambda}_n).  }

The classical equation of motion $E'(\phi)=0$ is a set of independent
algebraic equations, $V'(\lambda_n)=0$. So the solution at
$\theta=\infty$ is a sequence of components $({\lambda}_0,
{\lambda}_1, {\lambda}_2, ...)$ in the projection operator basis, with
each $\lambda_n$ being an extremum of the potential, and $\lambda_n=0$
as $n\rightarrow\infty$, which is the finite energy requirement for
the soliton solution. Here we assume the potential has zero vacuum
energy at the origin.

Therefore there are infinitely many soliton solutions at
$\theta=\infty$, as long as the potential has more than one
extremum. Each solution can be regarded as a map from the positive
integer to the extrema of the potential. A general solution will
spontaneously break the $U(\infty)$ group down to a finite unitary
subgroup, depending on how many $\lambda_n$ are the same in that
particular solution. This has been interpreted as describing the decay
of the unstable D-brane into multiple lower dimensional D-branes in
the string theory\refs{\hklm}.

\newsec{Noncommutative solitons at finite $\theta m^2$: general analysis}

In this section we will discuss the noncommutative soliton solutions
at finite $\theta$ in 2+1 dimensions. We assume the potential has a
true vacuum at the origin with value zero. This can always be
satisfied by a constant shift of the scalar field, if the highest
power term is even with positive coefficient. So for example, $\phi^3$
potential does not satisfy this condition while $\phi^4$ potential
does. This trivial looking assumption turns out to be essential in
understanding the existence of the critical point for the solitons. We
will comment on the case when the potential has odd highest power at
the end of the section.

The kinetic energy has to be taken into account. It breaks the
$U(\infty)$ symmetry, so the energy functional contains the unitary
matrix $U$. Let $U_{mn}=\langle m|U|n\rangle$ be the matrix element of
$U$ in SHO basis, the energy functional is
\eqn\energyD{ 
E(\{\lambda_n\},
\{U_{mn}\})=\sum_{n=0}^{\infty}\lambda_n^2(1+2\sum_{m=0}^{\infty}m|U_{mn}|^2)
-2\sum_{m,n=0}^{\infty}\lambda_m\lambda_n|A_{mn}|^2+\theta\sum_{n=0}^{\infty}V(\lambda_n),
}  where
\eqn\A{
A_{mn}=\sum_{k=1}^{\infty} \sqrt{k} U_{kn}U_{k-1,m}^* .}

Only the radially symmetric soliton solutions will be considered,
which means the scalar field in the operator representation is
diagonalized, $U=I$. Actually adding a noncommutative $U(1)$ gauge
field into the action can restore the $U(\infty)$ symmetry, while the
scalar field lies in the adjoint representation of this $U(\infty)$
group. Then by a proper $U(\infty)$ transformation, the radially
symmetric form of the scalar field can always be assumed.

Under such an assumption, the energy functional simplifies greatly,
\eqn\ReducedE{
E(\{\lambda_n\}) = \sum_{n=0}^{\infty}[(2n+1)\lambda_n^2 -
 2(n+1)\lambda_{n+1}\lambda_n + \theta V(\lambda_n)].  }   The
 classical equation of motion $\partial E /\partial \lambda_n =0$
 becomes a set of infinite number of coupled equations
\eqn\eomb{\eqalign{
  (n+1){\lambda}_{n+1} -(2n+1){\lambda}_n + n{\lambda}_{n-1} &=
  {1\over 2}
\theta V'({\lambda}_n), \qquad n\geq 1 \cr
\lambda_1-\lambda_0 & ={1\over
 2}  \theta V'({\lambda}_0). }}   In addition we impose the asymptotic
boundary condition required by the finiteness of the total energy
\eqn\BCone{ 
\lambda_n \rightarrow 0,\qquad {\rm as} \qquad n \rightarrow \infty 
}

This is a second order difference equation for which it is hard to
find a closed form solution. In general, the difference equation
allows more solutions than its corresponding differential equation.
In this section we first try a qualitative analysis to find the effect
of $\theta$ on the solution. In the next section, we will use both
numerical and analytical methods to analyze the $\phi^4$ potential in
detail.  m Add the set of equations \eomb\ up to the N-th to get an
equation that is ``integrated'' once
\eqn\firstintegration{
\lambda_{N+1}-\lambda_N={\theta \over 2} {1\over {N+1}} \sum_{n=0}^N V'(\lambda_n).
}  Take the $N\rightarrow\infty$ limit and use asymptotic condition
\BCone, we obtain the necessary condition
\eqn\constrainta{
\sum_{n=0}^{\infty}V'(\lambda_n)<\infty.
}

If we regard $n$ as the discrete time, the above equation describes a
non-autonomous dynamic system. The noncommutative soliton solution is
like a particle starting from a large nonzero $\lambda_0$ and
approaches zero as time $n$ goes to infinity. In general it is
possible for this particle to go back and forth, but it should
ultimately approach zero monotonically as dictated by the asymptotic
condition. We will only study  this part below, and previous
``motion''  only shows up as an initial condition of ${1\over
N}\sum_{n=0}^{N-1} V'(\lambda_n)$ which is bounded for $\lambda>0$.

First let us observe whether the particle approaches zero from below
or above. Assume this imaginary particle starts from positive
$\lambda$ and decreases. If at some time $N$, it jumps to
$\lambda_N<0$ close to zero, then $\sum_{n=0}^{N-1}
V'(\lambda_n)<0$. As the origin is a true vacuum, $V'$ is a line with
positive slope near the origin so $V'(\lambda_N)<0$. Then necessarily
$\lambda_{N+1}<\lambda_N<0$ from
\firstintegration\ and $\lambda_n$ would not converge to zero.
Similarly, if the particle starts from the negative point, it should
approach the origin from below. So we need only consider $\lambda_n$
positive only. Then for large $n$, $\lambda_n$ approach zero from
above monotonically, which by \firstintegration gives a sharper
constraint than
\constrainta
\eqn\constraintb{
\sum_{n=0}^{\infty}V'(\lambda_n) \leq 0.
}

\ifig\Poten{ The general $\phi^4$ potential $V(\phi)$ and its derivative $V'(\phi)$. Notice the absolute value of the minima (the valley) of $V'(\phi)$ for $\phi>0$ is smaller than the maxima of $V'(\phi)$ (the hump) in this case. 
}{\epsfxsize=4.0in \epsfbox{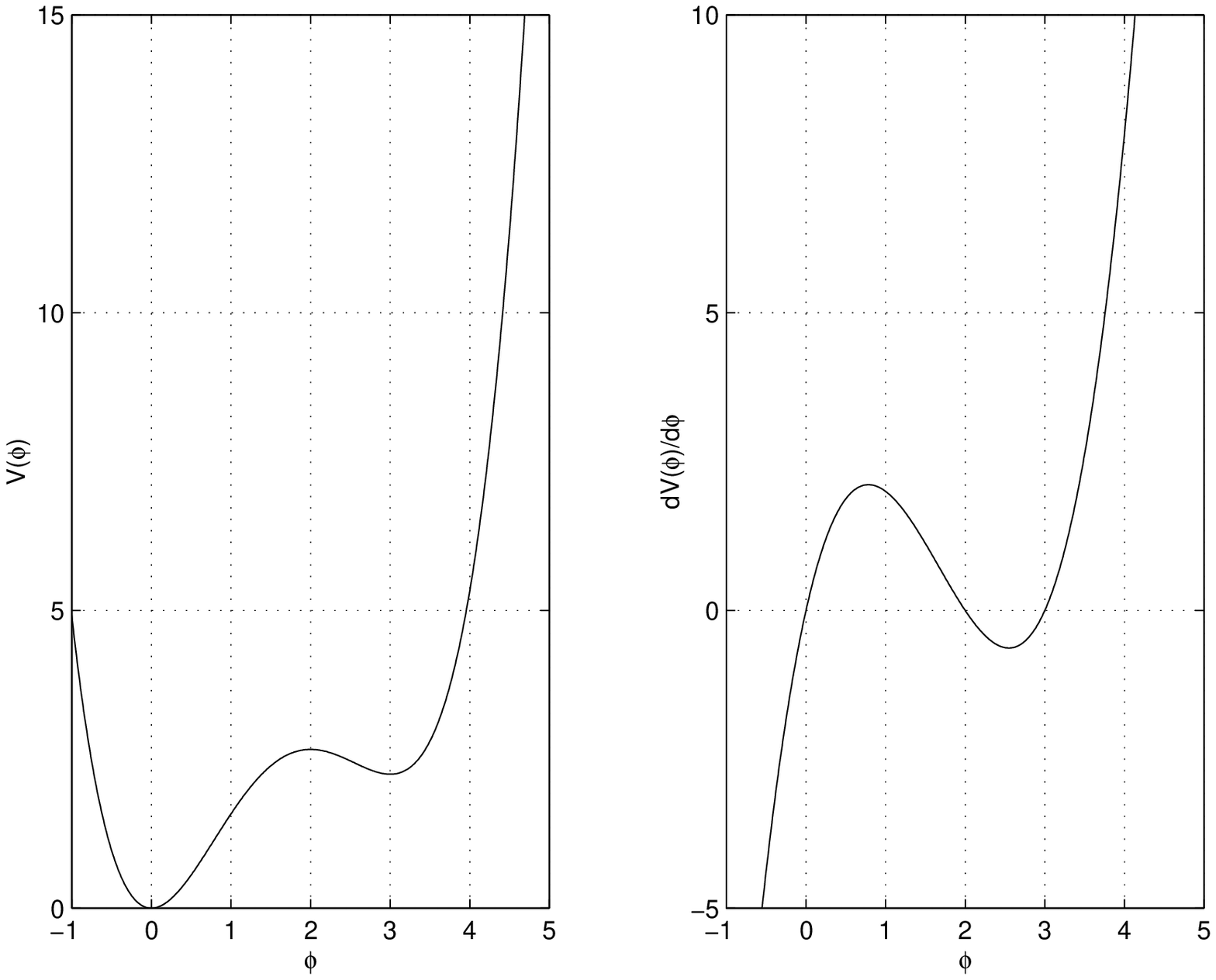}}

To see what happens as $\theta$ decreases, let us take a general
$\phi^4$ potential as an example, as shown in \Poten , and study the
extrapolation of the solution corresponding to the particle's starting
from a nonzero  $\lambda_0>0$ and jumping to $\lambda_1<<1$ and then
approach zero afterwards. This soliton solution has the lowest
$\theta_c$, as will be proved in the next section. \firstintegration\
requires $V'(\lambda_0)<0$ and $V'(\lambda_1)+V'(\lambda_0)<0$. But
the minimum of $V'$ is bounded for $\lambda>0$, so no matter how
$\lambda_0$ changes, $\lambda_1=\lambda_0+\theta V'(\lambda_0)/2$
would finally increase with $\theta$ decreasing when $\theta$ becomes
small enough, and the particle climbs up the hump at $\lambda_1$. In
other words, $\theta$ controls how far the imaginary particle can go
at each step. In this $\phi^4$ potential, it can be easily proved that
the requirement $V(x)\geq V(0)$ always makes the hump larger than the
absolute value of the valley. So as $\theta$ becomes small enough,
it's impossible for $\lambda_1$ to be close enough to the origin to
satisfy the necessary condition $V'(\lambda_1)+V'(\lambda_0)<0$. Then
$\lambda_2>\lambda_1$ and $\lambda_n$ wouldn't converge to zero. Thus
this solution can't exist for such a $\theta$.

It seems that the above argument depends on the special property of
the $\phi^4$ potential and presumably would not hold for a more
general polynomial potential. But there is a more general argument to
establish the existence of the finite critical point, although this
may not reflect the actual situation of how the solution
disappears. Notice that there is a bound on $\Delta \lambda_N\equiv
\lambda_N-\lambda_{N-1}$
\eqn\step{
|\Delta \lambda_N| = |{\theta \over 2} {1\over N} \sum_{n=0}^{N-1}
V'(\lambda_n)| <  {\theta \over 2} |{\rm inf}_{x>0}V'(x)|.     } When
$\theta$ becomes very small, $\Delta \lambda$ becomes very small, and
the sum can be approximated by an integral from $\lambda_0$ to
$\lambda_N$. This is exactly the same as using the limit process to
find Riemann integral of a function. But this integral
$V(\lambda_0)=\int_0^{\lambda_0}V'(x)dx> 0$ because zero is the true
vacuum of the potential at the origin and $\lambda_0$ cab not be a
minimum, so there must be a critical point $\theta_c$ such that the
sum becomes zero or even positive at some large N. Then the convergent
process stops and the solution disappears.

It is interesting to observe how the difference equation allows a
solution to avoid the constraint $\int_0^{\lambda_0}V'(x)dx>0$ in the
commutative limit and satisfies the constraint \constraintb. It is
comparable to the definition of the Riemann integral through the limit
of a finite sum by using a particular partition of the coordinate
region determined by $\lambda_0>\lambda_1>\lambda_2>...>0$ and letting
the partition be smaller and smaller. When $\theta$ is big enough, the
partition is coarse and it is possible for the sum to be different
from the continuous limit and remain always negative. Decreasing
$\theta$ is the same as taking the continuous limit to calculate the
Riemann integral and it will finally make the solution impossible at
some finite $\theta_c$.  In this sense, the noncommutative scale
$\theta$ really controls the continuum limit.

For the polynomial potential whose highest power is odd, the above
proof breaks down as either $V'(\phi)\rightarrow -\infty$ as $\phi
\rightarrow \infty$ or $V'(\phi)\rightarrow \infty$ as
$\phi\rightarrow -\infty$. So in principle, it is always possible to
find a soliton solution satisfying the required constrains at any
nonzero $\theta$. Certainly the commutative limit $\theta=0$
invalidates the scaling transformation of $x \rightarrow
x/\sqrt{\theta}$ and so it is a singular critical point of all the
soliton solutions in this case.

Our qualitative analysis is also valid for non-polynomial potentials,
such as the periodic cosine shaped potential, as long as it satisfies
the bounded-below condition.

\newsec{Noncommutative Solitons at finite $\theta$: $\phi^4$ potential}

In this section we discuss the $\phi^4$ potential in detail. First we
will use the numerical method to explicit construct these solutions at
finite $\theta$ and see clearly how the solutions varies with
$\theta$.  Second we will see that the method in \refs{\gms} can be
extended to find the lowest critical point $(\theta m^2)_c$ explicitly
to a very good approximation. Finally  using a simple scaling
argument, we find that $(\theta m^2)_c$ is the same for all the
symmetric $\phi^4$ potentials. Note, we use the dimensionless
paremeter $\theta m^2$ in this section instead of $\theta$, where
$m^2=V'(0)$.

\subsec{Numerical Results}

First we explain the numerical method briefly. We use the relaxation
method normally applied in solving differential equation with two
point boundary conditions. The two boundary conditions for \eomb\ are
\eqn\BCtwo{
{\lambda}_{-1}={\lambda}_{-2}=0,  \lambda_n \rightarrow 0 \ {\rm as} \
n\rightarrow \infty.} An initial guess for the solution is required as
an input. We can estimate the asympotic value of $\lambda_n$ by going
to the continuum limit and convert the difference equation into a
differential equation, and ignore the nonlinear terms
\eqn\eomc{
\lambda (u)={2\over {\theta m^2}}u{d^2\lambda (u) \over {du^2}.}
}  The solution satisfying the asympotic boundary condition is
\refs{\gms}
\eqn\Guess{
\lambda (n) = An^{1\over 4}e^{-\sqrt{2n\over {\theta m^2}}}
} which can be used as the initial input to recursively find the true
solution of the difference equation.

We use as an illustrative $\phi^4$ potential
\eqn\SpecialV{ 
V(\phi)={1\over 4}\phi^4-{5\over 3}\phi^3+3\phi^2 }  which is shown in
\Poten . It has a local maximum at $\phi =2$ and a local minimum at
$\phi =3$ in addition to the global minimum at $\phi=0$.

The results from the numerical analysis are as follows:

1) At $\theta=\infty$ and ignoring the kinetic term,
$\phi(x)=\sum_n\lambda_n\phi_n(x)$ with $V'(\lambda_n)=0$ is the
general soliton solution. There are several changes to the solution
after including the kinetic term at finite $\theta$. First, those
$\lambda_n$ which are zero at $\theta=\infty$ become nonzero. They
increase when $\theta$ decreases, but never become appreciably large.
Second, those $\lambda_i$ which are nonzero at $\theta=\infty$ will
also change such that $V'(\lambda_n)$ is negative and decreasing when
$\theta$ decreases. So if $\lambda_i$ at $\theta=\infty$ is a local
minimum/maximum, it will decrease/increase when $\theta$ decreases. A
typical example is the extrapolation of the solution
${\phi}=3{\phi}_0$ to $\theta m^2=600$, $\phi (x) =
2.89\phi_0+0.047\phi_1+0.0015\phi_2+0.00006\phi_3+...$. Here
$\lambda_0$ decreases with $\theta$ because $\lambda=3$ is the local
minimum of the illustrative potential given above. Finally, while
several $\lambda_n$ can take the same nonzero value at
$\theta=\infty$, this is not possible at finite $\theta$. For example,
the solution $\phi(x)=3\phi_1(x)+3\phi_2(x)$ at $\theta=\infty$,
extrapolated to $\theta m^2=78$, is
$\phi(x)=0.038\phi_0+2.90\phi_1+2.70\phi_2+0.10\phi_3+0.004\phi_4...$,
which clearly shows that $\lambda_1\neq\lambda_2$.

2) The existence of the critical point $\theta_c$ can be seen as
   follows. The solution at $\theta=\infty$ is characterized by those
   nonzero $\lambda_n$'s. If one of them becomes zero (approximately),
   then this particular solution can be regarded as being
   nonexistant. A nonzero $\lambda_n$ at $\theta=\infty$ always
   changes in the direction such that $V'(\lambda_i)$ decreases when
   $\theta$ decreases. Because $V'(\lambda)$ is bounded below for
   $\lambda\leq 0$, $\lambda_n$ will reach a critical value at a
   finite $\theta_c$. With further decrease in $\theta$, this
   particular $\lambda_n$ will jump to a small value which is
   approximately zero, and this solution will cease to exist.

3) When $\theta$ decreases, the nonzero $\lambda_n$ with the largest
   $n$ reaches the critical value first.  This can be explained
   considering the argument of section 3. The quantity $\Delta
   \lambda_N\equiv\lambda_N-\lambda_{N-1}$ should be negatively large
   enough to make the solution possible. As it is proportional to
   $1/N$, the $\lambda_N$ with the larger $N$ will fail this criterion
   first. So the solution $\phi=\lambda_0\phi_0$ gives the lowest
   critical point $\theta_c$.

4) We emphasize that at the critical point the solitonic solution has
   no singular behavior because $\theta_c$ is finite. The solution
   looks almost the same as the example given in 1). The only sign of
   the criticality is the discontiuity of the $\lambda_n$, which
   changes from a finite nonzero value to zero.

\subsec{Determination of $(\theta m^2)_c$}

In this section we will find the method of \refs{\gms} useful to
explicitly express the critical point $(\theta m^2)_c$  within a good
approximation. In particular for all the symmetric $\phi^4$ potential,
$(\theta m^2)_c$ is the same, which can be proved using a scaling
argument.

The lowest critical point $(\theta m^2)_c$  corresponds to the
extrapolation of the solution $\phi(x)=\lambda_0\phi_0(x)$. The
numerical solution indicates that $\lambda_n$ is extremely small for
$n\neq 0$, so ignoring the nonliear terms is a good approximation
\eqn\Linearized{
(n+1){\lambda}_{n+1}-(2n+1){\lambda}_n-n{\lambda}_{n-1} = {1\over
2}\theta m^2\phi . }  Going back to the coordinate space
representation, and noticing that $\lambda_0>>0$ is equivalent to
adding a source term proportional to $\phi_0(x)$
\eqn\PSpace{
(-{1\over{\theta m^2}}{\partial}^2+1)\phi(x)=A{\phi}_0(x), } except
that it has a different boundary condition at $n=0$
\eqn\BCthree{
{\lambda}_0={2 \over {\theta m^2}}({\lambda}_1-{\lambda}_0)+A . }
Compatibility with the boundary condition of the original equation of
motion determines A.

Equation \PSpace\ is solved using properties of Laguerre polynomials
${\phi}_n$,
\eqn\Expression{
\lambda_n = e\int_0^{+\infty}{e^{-x}\over {1+{2\over {\theta m^2}}x}}L_n(x) .
} Define the function F by
\eqn\coeff{
F(a) = \int_0^{+ \infty } {e^{-x}\over {1+2ax}}, } as shown in fig.2
, and $\lambda_0=F({1\over {\theta m^2}})$. The two boundary
conditions \BCthree\ and  \BCtwo\  should agree, which gives an
equation for the scale factor A
\eqn\BCthreeone{
m^2 A(F({1\over {\theta m^2}})-1) = V'(AF) = m^2(AF)-a(AF)^2+b(AF)^3,
} where we assume a general form of the $\phi^4$ potential,  $V'(\phi)
= m^2\phi-a\phi^2+b\phi^3$. The existence of a real solution for A
requires $({1\over {\theta m^2}}) \geq {4m^2b \over a^2}$. $F$ is a
monotonic function, so the equality determines the critical point
$(\theta m^2)_c$
\eqn\critpoint{
F({1\over {(\theta m^2)_c}}) = {4m^2b \over a^2}. } At this critical
point, ${\lambda}_0$ is equal to
\eqn\critical{
\lambda_c=(eF)_c={a\over {2b}.}
}  These two expressions agree with the numerical results.

\ifig\funcF{
Function F(a) \coeff\ which determines the coefficient
$\lambda_0=F({1\over {\theta m^2}})$ of the single $\delta$ function
like solution.  }{\epsfxsize=3.0in \epsfbox{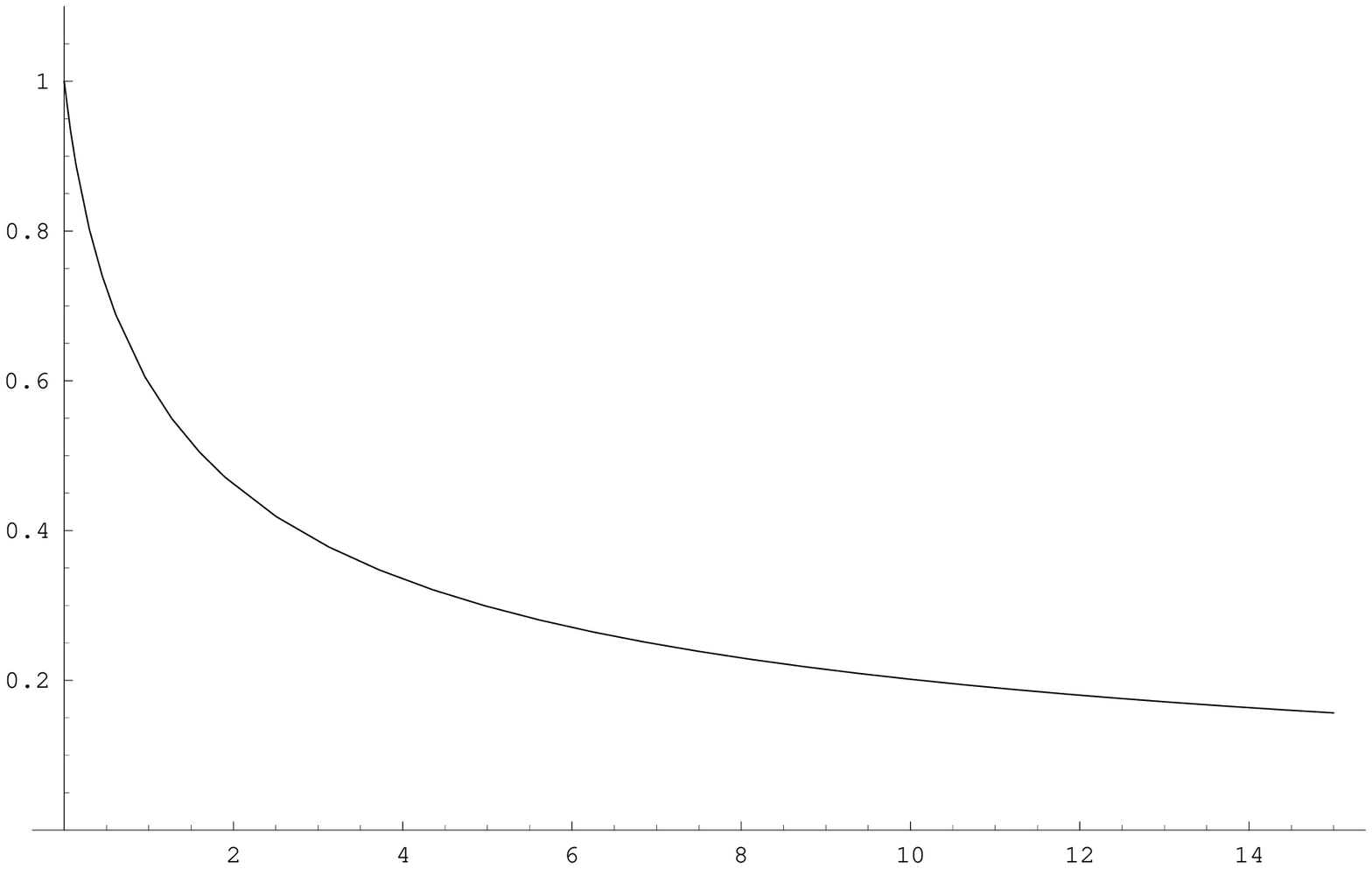}}

We have assumed that the potential has a global minimum at the origin,
so  $V(\phi) \geq 0$. It sets a lower bound $4m^2b/a^2 \geq 8/9$. By
\critpoint , it sets a lower bound for $F_c$, which in turn determines
a lower bound  $(\theta m^2)_c
\geq 14.374$. It is saturated exactly by the symmetric $\phi^4$
potential. Numerical analysis gives the exact result to be $(\theta
m^2)_c=13.92$.

Let's study the case of the symmetric $\phi^4$ potential in more
detail. The effect of the potential on $(\theta m^2)_c$ enters through
its derivative $V'$, as seen in the equation of motion in the
projection operator basis, equation \eomb. Assuming one of the
degenerate vacuum is at the origin, the symmetric $\phi^4$ potential
is characterized by the zeros of its derivative, assumed to be at 0,
1/a, 2/a. So in general $V'(\lambda)={m^2 \over
2}\lambda(a\lambda-1)(a\lambda-2)$, and the variation of $a$ and $m^2$
gives all the symmeric $\phi^4$ potential. Writing out \eomb\
explicitly as
\eqn\eomba{
  (n+1){\lambda}_{n+1}-(2n+1){\lambda}_n + n{\lambda}_{n-1} = {1\over
  2}\theta {m^2 \over 2}\lambda_n (a\lambda_n -1)(a\lambda_n -2). }
  Under scaling transformation $\lambda \rightarrow b\lambda$, it
  becomes
\eqn\eombb{
  (n+1){\lambda}_{n+1} -(2n+1){\lambda}_n + n{\lambda}_{n-1} = {1\over
  2}\theta {m^2 \over 2}\lambda_n (ab\lambda_n -1)(ab\lambda_n -2). }
  Effectively it transforms the moduli $a$ by the scaling factor $b$, which can be absorbed into $m^2$. This
  scaling of the variable will not affect the existence of the
  solution, and $\theta m^2$ remains invariant under the
  transformation, so the critical point $(\theta m^2)_c=13.92$ for the
  existence of the nontrivial solution is the same for all the
  symmetric $\phi^4$ potential.

\newsec{Discussion}

From the analysis in this paper, we find that noncommutative soliton
sector exhibits a nontrivial hierarchy structure controlled by the
noncommutative scale $\theta$. For quite general potentials which have
a global minimum, the $\theta_c$ is finite and determined by the
details of the potential. This indicates that the noncommutative
geometry should not be considered as merely a passive kinematic
background, but may have the similar dynamic content comparable to the
potential, as shown by its effect on the existence of the soliton
solutions.

The noncommutative soliton solutions can be interpreted as describing
the tachyon condensation on the unstable brane at large B field
background \refs{\dmr, \hklm}. It results in co-dimension-two branes
obtained from the decay of the original brane. A general
noncommutative soliton solution would disappear at finite
$\theta$. But notice that the string field theory algebra factors into
a  direct product only in the limit $B\rightarrow\infty$, and those
string degrees of freedom other than those involving only the center
of mass coordinates have to be taken into account at finite
$\theta$. This may change the picture in the string theory content.

The soliton solution at finite $\theta$ in general breaks the
$U(\infty)$  symmetry completely. At $\theta=\infty$ and ignoring the
kinetic energy, the level N soliton solution, such as $\phi =
\sum_{n=0}^N \lambda_n \phi_n$ with all $\lambda_n$ equal, break the
$U(\infty)$ down to $U(N)\times U(\infty-N)$ \refs{\hklm}. This
soliton solution describes N coincident D(n-2)-branes with $U(N)$
symmetry. Inclusion of the kinetic energy term brings $1/\theta^{-1}$
corrections and all the $\lambda_n$ are different. It breaks the
residual $U(N)$ symmetry completely into $U(1)^N$. So the $U(N)$
symmetry for this solution is at most approximate, and it seems that
it should be interpreted as describing N branes that are not
coincident. But this conclusion may not be true if we consider the
additional terms in the action coming from the open string field
theory, because at finite $\theta$ it is necessary to include those
string degree of freedom other than those accounted for by the
noncommutative field theory. In particular, noncommutative Yang-Mills
theory is not enough and the analysis of Dirac-Born-Infeld action may
give a different result which will be worth exploring.

Also notice that the open superstring tachyon potential for type II
superstring theory is exactly of the symmetric $\phi^4$ shape, which
follows from the reflection symmetry of the potential, although the
details of the potential are unknown. But we find the critical point
$(\theta m^2)_c=13.92$ for the existence of the noncommutative soliton
is ignorant of the exact form of the $\phi^4$ potential, and so is
charecteristic of type-II in the B-field background. We haste to add
that this conclusion may be changed after inclusion of the open string
field theory degrees of freedom.

It would be interesting to study the stability of these soliton
solutions, and their effects on the quantum structure of the whole
theory. We hope to explore these issues in the future.

{\bf Note added:}

We notice that the recent two papers \refs{\sochi, \gmsA} discussed
the noncommutative solitons in the context of noncommutative scalar
theory coupled with the noncommutative gauge field. These new
solutions involve nontrivial gauge field configutation, which may
answer some of the puzzles put forward in the last section of the
paper concerning the finite $\theta$ behavior and are worth exploring.

\centerline {\bf Acknowledgements}

The author cordially thanks Michael Douglas for invaluable advice,
discussions and encouragement, without which this work would not
have been possible.

\listrefs
\end